\documentclass{pnastwo}

\usepackage{amssymb,amsfonts,amsmath,graphicx}

\contributor{Submitted to Proceedings
of the National Academy of Sciences of the United States of America}
\url{www.pnas.org/cgi/doi/10.1073/pnas.0709640104}
\copyrightyear{2008}
\issuedate{Issue Date}
\volume{Volume}
\issuenumber{Issue Number}

\begin{document}



\title{On the unreasonable effectiveness of the post-Newtonian approximation in gravitational physics}





\author{Clifford M. Will\affil{1}{McDonnell Center for the Space Sciences, Department of Physics, Washington University, St. Louis, MO USA}}

\contributor{Submitted to Proceedings of the National Academy of Sciences
of the United States of America}

\maketitle

\begin{article}

\begin{abstract} 
The post-Newtonian approximation is a method for solving Einstein's field equations for physical systems in which motions are slow compared to the speed of light and where gravitational fields are weak.  Yet it has proven to be remarkably effective in describing certain strong-field, fast-motion systems, including binary pulsars containing dense neutron stars and binary black hole systems inspiraling toward a final merger.  The reasons for this effectiveness are largely unknown.  When carried to high orders in the post-Newtonian sequence, predictions for the gravitational-wave signal from inspiraling compact binaries will play a key role in gravitational-wave detection by laser-interferometric observatories.
 \end{abstract}






\section{Introduction}

The first detection of gravitational radiation, anticipated to occur during this decade, will undoubtedly be a triumph of experimental physics.  It will initiate a new kind of astronomy.  But it will also shine a spotlight on the theory of general relativity itself.  Gravitational radiation is a crucial prediction of Einstein's theory; indeed it is a natural prediction, given that the theory is built on a foundation of Lorentz invariance, which carries with it the concept of a limiting speed for interactions.  

On the other hand, general relativity is a notoriously complicated, non-linear, tensorial theory of the gravitational field.  Almost no physically useful exact solutions of the theory are known, and those that are known are endowed with such high degrees of symmetry that their realm of validity is limited.  To be sure, two of these solutions, due to Schwarzschild and Kerr, have proven to be of enormous importance, describing as they do the spacetime of isolated black holes, now widely accepted as being ubiquitous throughout the universe.  
Part of the Schwarzschild solution also describes the exterior geometry of any static spherical star or planet.  
Another extremely useful, albeit special solution is the Friedman-Robertson-Walker metric of the standard model of homogeneous and isotropic big-bang cosmology.  

But by its very nature, gravitational radiation involves spacetimes that are highly non-symmetrical and highly dynamical.  No exact solution of Einstein's equations is known that describes the emission and propagation of gravitational waves from a source, and the reaction of the source to the emission of those waves.

As a result, most of our understanding of gravitational radiation has come from {\em approximations} to Einstein's equations.  One class of approximations assumes that the gravitational fields in and around the source are suitably weak (the fields of the propagating waves weaken progressively as they leave the source), and that the motions within the source are suitably slow compared to the speed of light.  This class includes schemes known as post-Newtonian theory, which will be the main subject of this paper, and a related scheme known as post-Minkowki theory.  The underlying idea is to treat spacetime as being that of flat Minkowski spacetime as the zeroth approximation, and to modify it by successive corrections.

Another class of approximations takes a known exact solution of Einstein's equations, such as the black hole solutions of Schwarzschild or Kerr, and introduces small perturbations of those spacetimes, induced, for example, by a small particle orbiting the hole.   In cosmology, perturbations of the Friedman-Robertson-Walker solution permit treatment of the growth of large-scale structure in the universe and fluctuations in the cosmic background radiation.

A rather different approach to solving Einstein's equations is ``numerical relativity'', which endeavors to formulate and solve the exact equations to a precision limited only by available computer resources for highly dynamical, highly asymmetrical situations using accurate and robust numerical computations.   
In recent years, numerical relativity has significantly enhanced our understanding of colliding black holes and neutron stars and the associated emission of gravitational radiation.

Just as exact solutions of Einstein's equations have limited realms of validity, so too do approximation schemes.  Specifically, the post-Newtonian (PN) approximation is formally limited to weak gravitational fields and slow motions.  Yet recent experience has shown that the post-Newtonian approximation is ``unreasonably effective'' in describing physical systems whose characteristics either lie outside the technical realm of validity of the approximation or push up against the boundary of that realm.

The use of the term ``unreasonably effective'', and indeed the title of this paper, have been shamelessly appropriated from a famous 1960 paper by Eugene Wigner, entitled ``On the unreasonable effectiveness of mathematics in the physical sciences''~\cite{wigner}. 
In that paper, Wigner states `` $\dots$ the enormous usefulness of mathematics in the natural sciences is something bordering on the mysterious and $\dots$ there is no rational explanation for it''.   Einstein presented similar ideas in an address to the Prussian Academy of Sciences in 1921.

While the considerations of Einstein and Wigner deal with deep questions of the nature of knowledge of the physical world and how it is acquired and assembled, the topic of this paper is much more narrow.  Our purpose will be to illustrate the various ways in which the post-Newtonian approximation has proven its extraordinary effectiveness in gravitational physics.  Nevertheless it is no less mysterious: we have no good understanding of why this approximation to general relativity should be so effective.

Because most of these applications of post-Newtonian theory involve the motions of gravitating bodies in their mutual gravitational fields and the emission of gravitational radiation, we will begin with a history of the ``problem of motion'', one of the central challenges in the development of general relativity.  The problem has at times been contentious.  We will then describe briefly the nature and structure of the post-Newtonian approximation, and will review some of the recent developments in the subject, particularly those involving going to very high orders in the approximation.   We will describe the usefulness of the PN approximation for characterizing alternative theories of gravity, and as a tool in experimental gravitation.  Then we show how the PN approximation effectively and accurately accounts for the observed behavior of binary pulsar systems, despite the presence of strong gravitational fields in the interiors of the orbiting neutron stars.  Finally, we will describe how the PN approximation, when carried to high enough orders, effectively describes the motion of and gravitational radiation from
inspiralling binary black hole systems, well into the strong-field region, and how its predictions merge smoothly with those from numerical relativity.   We will conclude by returning to the mystery of this effectiveness.

\section{Motion and radiation in general relativity: A history}
\label{sec:motion}

At the most primitive level, the problem of motion in general relativity is relatively straightforward, and was an integral part of the theory as proposed by Einstein\footnote{This history will by necessity be personal and selective.  For a detailed technical and historical review of the problem of motion, see Damour~\cite{damour}}.  A ``test particle'', that is a particle whose mass is sufficiently small that its own contribution to the curvature of spacetime can be ignored, moves on a geodesic of the curved spacetime in which it finds itself.  Underlying this concept is the ``weak equivalence principle'', which states that the acceleration of a suitably small body in an external gravitational field is independent of its internal structure or composition, a principle verified by modern experiments to parts in $10^{13}$.  
Using the geodesic equation and an approximate solution for the spacetime metric around the Sun, Einstein was able in 1915 to obtain the relativistic contribution to the perihelion advance of Mercury, in agreement with observations.   

The first attempts to treat the motion of multiple bodies, each with a finite mass, were made in the period 1916--1917 by de Sitter, Lorentz and Droste~\cite{desitter,lorentz}.  They derived the metric and   equations of motion for a system of $N$ bodies, in what today would be called the first post-Newtonian approximation of general relativity (de Sitter's equations turned out to contain some important errors).   In 1916, Einstein took the first crack at a study of gravitational radiation, deriving the energy emitted by a body such as a rotating rod or dumbbell, held together by non-gravitational forces~\cite{einstein}.  He made some unjustified assumptions as well as a trivial numerical error (later corrected by Eddington~\cite{eddington}), but the underlying conclusion that dynamical systems would radiate gravitational waves was correct.  

The next significant advance in the problem of motion came 20 years later.  In 1938, Einstein, Infeld and Hoffman published the now legendary ``EIH'' paper, a calculation of the $N$-body equations of motion using only the vacuum field equations of general relativity~\cite{EIH}.  They treated each body in the system as a spherically symmetric object whose nearby vacuum exterior geometry approximated that of the Schwarzschild metric of a static spherical star.   They then solved the vacuum field equations for the metric between each body in the system in a weak field, slow-motion approximation.  Then, using a primitive version of what today would be called ``matched asymptotic expansions'' they showed that, in order for the nearby metric of each body to match smoothly to the interbody metric at each order in the expansion, certain conditions on the motion of each body had to be met.  Together, these conditions turned out to be equivalent to the Droste-Lorentz $N$-body equations of motion.  Remarkably, the internal structure of each body was irrelevant, apart from the requirement that its nearby field be approximately spherically symmetric.

Around the same time, there occurred an unusual detour in the problem of motion.  Using equations of motion based on de Sitter's paper, specialized to two bodies, 
Levi-Civita~\cite{levicivita} showed that the center of mass of a binary star system would suffer an acceleration in the direction of the pericenter of the orbit, in an amount proportional to the difference between the two masses, and to the eccentricity of the orbit.  Such an effect would be a violation of the conservation of momentum for isolated systems caused by relativistic gravitational effects.  Levi-Civita even went so far as to suggest looking for this effect in selected nearby close binary star systems.  However,  Eddington and Clark~\cite{eddingtonclark} quickly pointed out that Levi-Civita had based his calculations on de Sitter's flawed work; when correct two-body equations of motion were used, the effect vanished, and momentum conservation was upheld.  Robertson confirmed this using the EIH equations of motion~\cite{robertson}.  Ironically, the acceleration of the center of mass of a binary system is today a subject of great theoretical and astrophysical interest, albeit for a very different reason.  We will return to this subject later.

Roughly 20 more years would pass before another major attack  on the problem of motion.  This period was the continuation of a time of relative dormancy for the entire subject of general relativity that lasted from the 1920s until the 1960s.  This dormancy resulted in part from the lack of experimental or observational relevance for the theory, in part from the perceived  complexity of the theory, and in part from the emergence of new fields of physics such as nuclear and particle physics in the middle part of the 20th century.

But in the middle 1960s, when a revival of general relativity was in its early phase, Fock in the USSR and Chandrasekhar in the US independently developed and systematized the post-Newtonian approximation in a form that laid the foundation for modern post-Newtonian theory~\cite{fock,chandra}.  They developed a full post-Newtonian hydrodynamics, with the ability to treat realistic, self-gravitating bodies of fluid, such as stars and planets.  In the suitable limit of ``point'' particles, or bodies whose size is small enough compared to the interbody separations that finite-size effects such as spin and tidal interactions can be ignored, their equations of motion could be shown to be equivalent to the EIH and the Droste-Lorentz equations of motion.  Chandrasekhar and his students also began extending the theory to higher orders in the post-Newtonian approximation.

An important byproduct of the Fock-Chandrasekhar work was the discovery by Nordtvedt~\cite{nordtvedt} that, in theories of gravity alternative to general relativity, the motion of self-gravitating bodies {\em could} depend on their internal structure, in contrast to general relativity, where a body's internal structure is irrelevant to its motion (barring tidal or spin effects).  
As a result of this ``Nordtvedt effect'', the Earth and the Moon could fall toward the Sun with slightly different accelerations, because of the small difference in their internal gravitational binding energy per unit mass.
Nordtvedt's discovery led to an important new test of general relativity using Lunar laser ranging, and to the development by Nordtvedt and by Will of the parametrized post-Newtonian (PPN) framework for treating alternative theories and experimental tests.  This will be the subject of a later section.

The next important period in the history of the problem of motion was 1974 -1979, initiated by the 1974 discovery of the binary pulsar PSR 1913+16 by Hulse and Taylor~\cite{hulsetaylor}.  Around the same time there occurred the first serious attempt to calculate the head-on collision of two black holes using purely numerical solutions of Einstein's equations, by Smarr and collaborators~\cite{smarr}, building on the pioneering work by Hahn and Lindquist~\cite{hahn}  

The binary pulsar consists of two neutron stars, one an active pulsar detectable by radio telescopes, the other very likely an old, inactive pulsar.  Each neutron star has a mass of around 1.4 solar masses.  The orbit of the system was seen immediately to be quite relativistic, with an orbital period of only eight hours, and a mean orbital speed of 200 km/s, some four times faster than Mercury in its orbit.  Within weeks of its discovery, numerous authors pointed out that PSR 1913+16 would be an important new testing ground for general relativity.  In particular, it could provide for the first time a test of the effects of the emission of gravitational radiation on the orbit of the system.  

However, the discovery revealed an ugly truth about the ``problem of motion''.  As Ehlers {\em et al.} pointed out in an influential 1976 paper~\cite{ehr}, the general relativistic problem of motion and radiation was full of holes large enough to drive trucks through.  They pointed out that most treatments of the problem used ``delta functions'' as a way to approximate the bodies in the system as point masses.   As a consequence, the ``self-field'', the gravitational field of the body evaluated at its own location, becomes infinite.   While this is not a major issue in Newtonian gravity or classical electrodynamics, the non-linear nature of general relativity requires that this infinite self-field contribute to gravity.   In the past, such infinities had been simply swept under the rug.   Similarly, because gravitational energy itself produces gravity it thus acts as a source throughout spacetime.  This means that, when calculating radiative fields, integrals for the multipole moments of the source that are so useful in treating radiation begin to diverge.  These divergent integrals had also been routinely swept under the rug.  Ehlers {\em et al.} further pointed out that the true boundary condition for any problem involving radiation by an isolated system should be one of ``no incoming radiation'' from the past.  Connecting this boundary condition with the routine use of retarded solutions of wave equations was not a trivial matter in general relativity. 
Finally they pointed out that there was no evidence that the post-Newtonian approximation, so central to the problem of motion, was a convergent or even asymptotic sequence.  
Nor had the approximation been carried out to high enough order to make credible error estimates.

During this time, some authors even argued that the ``quadrupole formula'' for the gravitational energy emitted by a system (see below), while correct for a rotating dumbell as calculated by Einstein, was actually {\em wrong} for a binary system moving under its own gravity.  The discovery in 1979 that the rate of decay of the orbit of the binary pulsar was in agreement with the standard quadrupole formula made some of these arguments moot.  Yet the question raised by Ehlers {\em et al.} was still relevant: is the quadrupole formula for binary systems an actual prediction of general relativity?

Motivated by the Ehlers {\em et al.} critique, numerous workers began to address the holes in the problem of motion, and by the late 1990s most of the criticisms had been answered, particularly those related to divergences.  The one question that remains open is the nature of the post-Newtonian sequence; we still do not know if it converges, diverges or is asymptotic.  Despite this, it has proven to be remarkably effective.

The final important development in this history of the problem of motion was the proposal to build large-scale laser interferometric gravitational-wave observatories, both on the ground and in space~\cite{ligo,virgo,lisa}.
It was quickly realized that, in order to maximize the likelihood of detecting the leading candidate source of gravitational waves -- the final inspiral and merger of binary systems of neutron stars or black holes, {\em extremely} accurate theoretical predictions of the gravitational waveform signal would be needed.  This meant that calculations of the equations of motion and gravitational waves from binary systems would have to be carried out to many orders beyond the simple post-Newtonian approximation.   The completion of this ambitious program by many groups worldwide has led to the discovery of the unreasonable effectiveness of post-Newtonian theory in the extreme realm of the merger of compact astronomical bodies.

\section{The post-Newtonian approximation}
\label{sec:PN}

The post-Newtonian approximation is based on the assumption that gravitational fields inside and around bodies are weak  and that characteristic motions of matter are slow compared to the speed of light.  This means that one can characterize the system in question by a small parameter $\epsilon$, where
\begin{equation}
\epsilon \sim (v/c)^2 \sim GM/rc^2 \sim p/\rho c^2 \,,
\end{equation}
where $v$, $M$ and $r$ denote the characteristic velocity, mass, and size or separation within the system; $p$ and $\rho$ are the characteristic pressure and density within the bodies; $G$ and $c$ are Newton's gravitational constant and the speed of light, respectively.

One then incorporates this approximation into methods for solving Einstein's equations.
Those equations, $G_{\mu\nu} = 8\pi (G/c^4 )T_{\mu\nu}$ are elegant and
deceptively simple, showing how geometry (in the form of the Einstein
tensor $G_{\mu\nu}$, which is a function of spacetime curvature)
is generated by matter (in the form of the material energy-momentum tensor
$T_{\mu\nu}$).  However, this is not the most useful form for actual
calculations.  For post-Newtonian calculations, a far more useful form
is the set of so-called ``relaxed'' Einstein equations:
\begin{equation}
\Box h^{ \alpha \beta } = -16 \pi (G/c^4 ){\tau}^{ \alpha \beta },
\label{relaxed}
\end{equation}
where $\Box \equiv  -{\partial}^2 / \partial (ct)^2 + {\nabla}^2 $
is the flat-spacetime wave operator,
$h^{ \alpha \beta }$ is a ``gravitational tensor potential''
related to the deviation of the spacetime metric $g_{\alpha\beta}$ from its flat-spacetime Minkowski
form $\eta_{\alpha\beta}$ by the formula
$h^{\alpha \beta} \equiv \eta^{\alpha \beta} - (-g)^{1/2} g^{\alpha
\beta}$, where $g$ is the determinant of $g_{\alpha
\beta}$, and where a particular coordinate system has been specified
by the de~Donder
or harmonic gauge condition
$\partial h^{\alpha \beta} /\partial x^\beta =0$ (summation on
repeated indices is assumed, with $x^0=ct$).  Because we assume that gravity is weak everywhere, the field $h^{\alpha \beta}$ is ``small''.
This form of Einstein's equations bears a striking similarity to Maxwell's
equations for the vector potential $A^\alpha$ in Lorentz gauge: $\Box
A^\alpha = -4\pi J^\alpha$, $\partial A^\alpha /\partial
x^\alpha =0$.   The key difference is that the source on the right
hand side of Eq.~(\ref{relaxed}) is given by the ``effective''
energy-momentum pseudotensor,
\begin{equation}
\tau^{\alpha\beta} = (-g)T^{\alpha\beta} + (16\pi)^{-1}
\Lambda^{\alpha\beta},
\label{effective}
\end{equation}
where $\Lambda^{\alpha\beta}$ is the non-linear ``field'' contribution
given by terms quadratic (and higher) in $h^{\alpha \beta}$ and its
derivatives (see~\cite{mtw}, Eqs.~(20.20, 20.21) for formulae).
In general relativity, the gravitational field itself is a source of
gravity, a reflection of the nonlinearity of Einstein's equations, and
in
contrast to the linearity of Maxwell's equations.

Equation~(\ref{relaxed}) is an exact restatement of Einstein's equations, and depends only on the assumption
that spacetime can be covered by harmonic coordinates.  It is called
``relaxed'' because it
can be solved formally as a functional of source variables without
specifying the motion of the source, in the form
\begin{equation}
h^{\alpha \beta} (t,{\bf x}) =  \frac{4G}{c^4} \int_{\cal C}
{ \tau^{\alpha \beta} (t -| {\bf x} - {\bf x'} |/c, {\bf x'}
)
\over | {\bf x} - {\bf x'} | } d^3x',
\label{nearintegral}
\end{equation}
where the integration is over the past flat-spacetime null cone $\cal
C$ of the field point $(t,{\bf x})$.
The motion of the source is then determined either by the equation
$\partial {\tau}^{\alpha \beta} /\partial x^\beta =0$ (which follows
from the harmonic gauge condition), or from the usual covariant
equation of motion $\nabla_\beta {T^{\alpha\beta}}=0$, where 
$\nabla_\beta$ denotes a covariant derivative.
This formal solution can then be iterated in a 
weak-field ($||h^{\alpha \beta}|| \ll 1$) approximation.  One begins by
substituting
$h_0^{\alpha \beta} =0$ into the source $\tau^{\alpha \beta}$ in Eq.
(\ref{nearintegral}), and
solving for the first iterate $h_1^{\alpha \beta}$, and then repeating the
procedure sufficiently many times to achieve a solution of the desired
accuracy.  This procedure is often called post-Minkowski theory.  If one further imposes the slow motion ($v \ll c$) assumption, one obtains post-Newtonian theory.  For example, to obtain the equations of motion at first post-Newtonian (1PN) order, {\it
two} iterations are needed (i.e.\ $h_2^{\alpha \beta}$ must be
calculated).   To obtain the leading gravitational waveform and energy flux far from
a binary system, two iterations are needed, while to obtain the leading contributions to gravitational radiation damping of the source, {\em three} iterations are necessary (see~\cite{walkerwill} for a discussion).  


However, because the source ${\tau}^{\alpha \beta}$ contains
$h^{\alpha \beta}$ itself, it is not confined to a compact region, but
extends over all spacetime.  As a result, as emphasized by Ehlers {\em et al.}~\cite{ehr}, there is a danger that the
integrals involved in various expansions of the solutions for $h^{\alpha \beta}$ will diverge or be
ill-defined.   Numerous approaches were developed to 
handle this difficulty.  The ``post-Minkowski'' method of Blanchet,
Damour and Iyer
solves Einstein's equations by two
different techniques, one in the near zone (within one gravitational wavelength of the source) and one in the far zone,
and uses the method of singular asymptotic matching to join the
solutions in an overlap region.  The method provides a natural
``regularization'' technique to control potentially divergent
integrals (see~\cite{BlanchetLRR} for a thorough review of this method).  
The ``Direct Integration of the Relaxed Einstein
Equations'' (DIRE) approach of Will, Wiseman and Pati (see~\cite{patiwill1}) retains
Eq.~(\ref{nearintegral}) as the global solution, but splits the
integration into one over the near zone and another over the far zone,
and uses different
integration variables to carry out the explicit integrals over the two
zones.  In the DIRE method, all integrals are finite and convergent.

The problem of ``delta-functions'' was handled in a variety of ways.
One was to import from quantum field theory a set of powerful techniques called ``dimensional regularization'', that, subject to some benign assumptions about analyticity, could be used to control the potential infinities order by order in the PN expansion.  Another adapted the EIH matching method to higher-order PN calculations.  A third approach treated each body as a real, nearly spherical fluid ball, and sorted contributions depending on how they scaled with the size of the ball.  In every case where different methods made a prediction about the equations of motion or the gravitational-wave signal, they were in complete agreement.




Among the results of these approaches are formulae for the equations of
motion and gravitational waveform of binary systems of compact
objects, carried out to high orders in a PN expansion.  Here we shall
only state a few formulae for the purpose of illustration.
For example,
the relative two-body equation of motion has the form
\begin{eqnarray}
\frac{d{\bf v}}{dt} &=& \frac{Gm}{r^2} \biggl\{- {\bf {\hat
      n}} +\frac{1}{c^2} {\bf A}_{1\rm PN} + \frac{1}{c^4}{\bf
A}_{2\rm PN} +\frac{1}{c^5} {\bf A}_{2.5\rm PN} 
\nonumber \\
&& \qquad \qquad+ \frac{1}{c^6}{\bf A}_{3\rm PN} + \frac{1}{c^7} {\bf
A}_{3.5\rm PN} + \dots
\biggr\},
\label{EOM}
\end{eqnarray}
where $m=m_1+m_2$ is the total mass, $r= |{\bf x}_1 -{\bf x}_2|$,
${\bf v}={\bf v}_1-{\bf v}_2$, and
${\bf {\hat n}} = ({\bf x}_1 -{\bf x}_2)/r$.
The notation ${\bf A}_{n\rm PN}$ indicates that the term is
$O(\epsilon^n)$ relative to the leading Newtonian term $-{\bf {\hat n}}$.
Explicit and unambiguous
formulae for non-spinning bodies through 3.5PN order have been
calculated by
various authors, and a number of spin-orbit and spin-spin contributions have been obtained (see~\cite{BlanchetLRR} for a review).
Here we quote only the first PN correction and the
leading radiation-reaction term at 2.5PN order:
{\setlength{\arraycolsep}{0.14 em}
\begin{eqnarray}
 {\bf A}_{1\rm PN} &=& \left\{ (4+2\eta)\frac{Gm}{r} - (1+3\eta)v^2 +
\frac{3}{2} \eta {\dot r}^2 \right\}{\bf {\hat n}} 
\nonumber \\
&& \qquad+ (4-2\eta) \dot r
{\bf v} \,, 
\label{APN} \\
{\bf A}_{2.5\rm PN} &=& - \frac{8}{15} \eta \frac{Gm}{r} \biggl\{ \left( 9v^2 +17 \frac{Gm}{r} \right) \dot r {\bf {\hat n}} 
\nonumber \\
&&  \qquad - \left( 3v^2 +9 \frac{Gm}{r}
\right) {\bf v} \biggr\} \,, 
\label{A2.5PN}
\end{eqnarray}}
where $\eta = m_1m_2/(m_1+m_2)^2$ and $\dot{r} = dr/dt$.
These terms are sufficient to analyse the orbit and evolution of
binary pulsars.  For example, the 1PN terms are
responsible for
the periastron advance of an eccentric orbit, given by $d \omega/dt =
6\pi f_{\rm b} Gm/[ac^2(1-e^2)]$, where $a$ and $e$ are the semi-major axis and
eccentricity, respectively, of the orbit, and $f_{\rm b}$ is the orbital
frequency, given to the needed order by Kepler's third law
$2 \pi f_{\rm b} = (Gm/a^3)^{1/2}$.

Another product is a formula for the gravitational field
far from the system, whose spatial components $h^{ij}$ (often called the gravitational ``waveform'') are sufficient to determine the signal detected by a laser interferometer, written schematically in the form
\begin{eqnarray}
h^{ij}(t, {\bf x}) &=& \frac{2Gm}{R c^4} \biggl\{ Q^{ij} + \frac{1}{c} Q_{0.5\rm PN}^{ij} +
\frac{1}{c^2} Q_{1\rm PN}^{ij} + 
\frac{1}{c^3}Q_{1.5\rm PN}^{ij} 
\nonumber \\
&& \qquad 
+ \frac{1}{c^4}Q_{2\rm PN}^{ij} + 
\frac{1}{c^5}Q_{2.5\rm PN}^{ij} +
\dots \biggr\} \,,
\label{waveform}
\end{eqnarray}
where $R$ is the distance from the source, and the variables
are to be evaluated at retarded time $t-R/c$.  The leading term
is the so-called quadrupole formula, given explicitly by
\begin{equation}
h^{ij}(t,{\bf x}) =  \frac{2G}{Rc^4}{\ddot I}^{ij}(t-R/c) \,,
\label{waveformquad}
\end{equation}
where $I^{ij}$ is the quadrupole moment of the source, and overdots
denote time derivatives.  For a binary system this leads to
$Q^{ij} = 2\eta (v^iv^j - Gm{\hat n}^i{\hat n}^j/r)$.
For binary systems, explicit
formulae for the waveform through 3.5PN order have been derived 
(see~\cite{BlanchetLRR} for a full review).

Given the gravitational waveform, one can
compute the rate at which energy is carried off by the radiation.
The lowest-order quadrupole formula leads to the
gravitational wave energy flux
\begin{equation}
\dot E = \frac{8}{15} \eta^2 \frac{G^2m^4}{r^4 c^5} (12v^2-11 {\dot r}^2).
\label{EdotGR}
\end{equation}
This has been extended to 3.5PN order beyond the quadrupole formula~\cite{BlanchetLRR}.
Formulae for fluxes of angular and linear momentum can also be
derived.
The 2.5PN radiation-reaction terms in the equation of motion~(\ref{EOM})
result in
a decrease of the orbital energy at a rate that precisely balances the energy
flux~(\ref{EdotGR})
determined from the waveform.  Averaged over one orbit, this
results in a rate of increase of the binary's orbital frequency  given by
\begin{eqnarray}
\frac{d f_{\rm b}}{dt}&=& \frac{192\pi}{5} f_{\rm b}^2 \left ( \frac{2\pi G{\cal M}f_{\rm b}}{c^3} \right )^{5/3} F(e),
\nonumber \\
F(e)&=&(1-e^2)^{-7/2}\left ( 1+ \frac{73}{24}e^2+\frac{37}{96} e^4 \right ) \,,
\label{fdotGR}
\end{eqnarray}
where ${\cal M}$ is the so-called ``chirp'' mass, given by ${\cal
M}=\eta^{3/5} m$.
Notice that by making precise measurements of the phase $\Phi (t) = 2\pi
\int^t
f(t') dt'$ of either the orbit or the gravitational waves
(for which $f =2f_{\rm b}$ for the dominant component) as a function of
the frequency, one in effect measures the ``chirp'' mass of the
system.

\section{The parametrized post-Newtonian framework and tests of general relativity}
\label{sec:PPN}

The post-Newtonian approximation has been remarkably effective as a tool for interpreting experimental tests of general relativity. 
This is because, in a broad class of alternative metric
theories of gravity, it turns out that only the 
values of a set of numerical coefficients in the post-Newtonian expression for the  spacetime metric vary from theory to theory.  Thus one can encompass a wide range of alternative theories by simply introducing arbitrary parameters in place of the numerical coefficients.  This idea dates
back
to Eddington in 1922, but the ``parametrized post-Newtonian (PPN) framework''  was fully developed by Nordtvedt and by Will in
the period 1968--72~\cite{nordtvedt68,will71,willnordtvedt}.  The framework
contains ten PPN parameters: $\gamma$, related to the amount
of spatial curvature generated by mass; $\beta$, related to
the degree of non-linearity in the gravitational field; $\xi$,
$\alpha_1$, $\alpha_2$, and $\alpha_3$, which determine whether the theory predicts that local gravitational experiments could yield results that depend on the location or velocity of the reference frame; and $\zeta_1$, $\zeta_2$, $\zeta_3$ and
$\zeta_4$, which describe whether the theory 
has appropriate momentum conservation laws.  In general relativity, $\gamma=1$, $\beta=1$, and the
remaining parameters all vanish.  For a complete exposition
of the PPN framework see~\cite{tegp}.

To illustrate the use of these PPN parameters in experimental tests, we cite the deflection of light by the Sun, an experiment that made Einstein an international celebrity when the sensational news of the Eddington-Crommelin eclipse measurements was relayed in November 1919 to a war-weary world.  For a light ray which passes a distance $d$ from the Sun, the deflection is given by
\begin{eqnarray}
\Delta \theta &=& \left (\frac{1+\gamma}{2} \right ) \frac{4GM}{dc^2}
\,, 
\end{eqnarray}
where 
$M$ is the mass of the Sun.   The ``$1/2$'' part of the coefficient can be derived by considering the Newtonian deflection of a particle passing by the Sun, in the limit where the particle's velocity approaches $c$; this was first calculated independently by Henry Cavendish and Johann von Soldner around 1800 (see, e.g.,~\cite{willajp}).   The second ``$\gamma/2$'' part comes from the bending of ``straight'' lines near the Sun relative to lines far from the Sun, as a consequence of the curvature of space.   
A related effect called the 
Shapiro time delay, an excess delay in travel time for light signals passing by the Sun, also depends on the coefficient $(1+\gamma)/2$.

Measurements using visible light, made during solar eclipses,
began with the 1919 measurements of
Eddington and his colleagues, but never reached better than ten per cent precision, largely because of the logistical difficulties inherent in such measurements.  High precision measurements were achieved using radio waves beginning in the late 1960s, culminating in
the use of Very Long Baseline Radio Interferometry (VLBI).   A  2004 analysis of VLBI data on
541 quasars and compact radio galaxies distributed over the entire
sky verified general relativity at the 0.02 percent level~\cite{shapiro04}.
Shapiro time delay
measurements began also in the late 1960s, by bouncing radar signals off Venus and Mercury or by tracking interplanetary spacecraft; the most recent test used tracking data from the {\em Cassini} spacecraft while it was en route to Saturn,
yielding a result at the 0.001
percent
level~\cite{bertotti}.  For a comprehensive review of the current status of experimental tests of GR, see~\cite{livrevCMW}.

Other experimental bounds on the PPN parameters came from measurements of the 
perihelion-shift of Mercury, searches for the ``Nordtvedt effect'' in the Earth-Moon orbit using
Lunar laser ranging, and a variety of
geophysical and astronomical observations.  All bounds were consistent with the predictions of general relativity, as
summarized in Table 1.

\section{Binary pulsars and the Strong Equivalence Principle}
\label{sec:binarypulsars}

Binary pulsars, such as the famous Hulse-Taylor system PSR 1913+16~\cite{hulsetaylor}, illustrate the unreasonable effectiveness of the post-Newtonian approximation.  

\begin{figure}[t]
\begin{center}
\includegraphics[width=7cm]{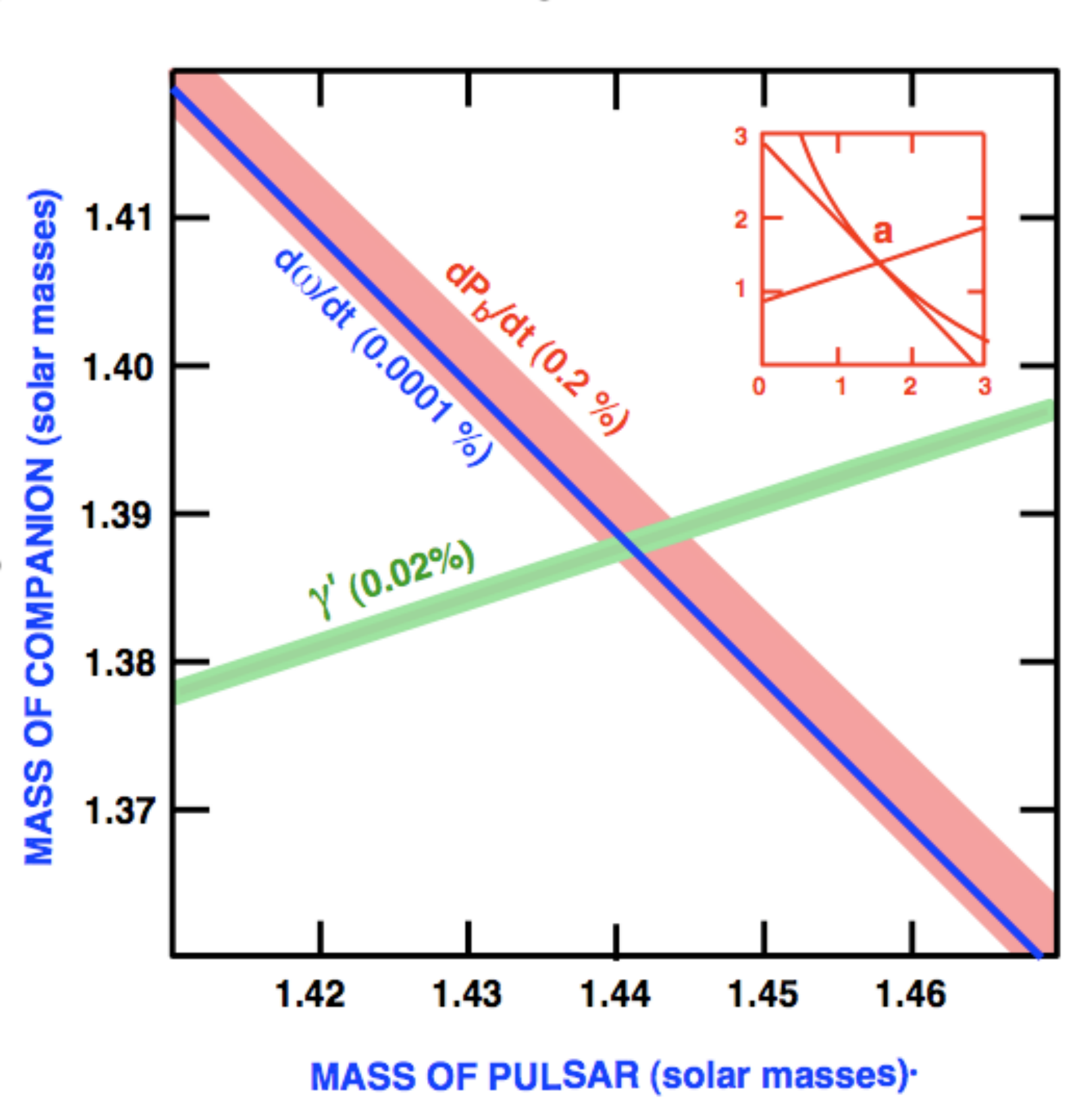}
\caption{Concordance between observations and the post-Newtonian predictions of general relativity for the binary pulsar PSR 1913+16.  Inset shows the full $m_1-m_2$ plane and the intersection region `a'.  The width of each band reflects the error in measuring each parameter.    
}
\label{Fig1}
\end{center}
\end{figure}

Through precise
timing of the pulsar ``clock'', the orbital parameters of the
system can be measured with exquisite precision.  These include
non-relativistic ``Keplerian'' parameters, such as the orbital eccentricity
$e$,
and the orbital period $P_b$, as well as a set of 
relativistic, or
``post-Keplerian'' parameters.   
The latter parameters include
the mean rate of
advance of periastron ($d\omega/dt$), the analogue of Mercury's perihelion shift;
the effect of 
special relativistic time-dilation and the gravitational redshift
on the observed phase or arrival time 
of pulses, resulting from the pulsar's orbital
motion and the gravitational potential of its companion (represented by a parameter $\gamma'$);
the rate of decrease of the orbital period ($dP_b/dt$),
taken to be the 
result of gravitational radiation damping (apart from a small
correction due to galactic differential rotation); and two
parameters related to the Shapiro time delay of the pulsar signal
as it passes by the companion.  According to general relativity,
the post-Keplerian effects depend only on $e$ and $P_b$, which are
known, and on the two stellar masses, which are unknown {\em a priori}.  By combining
the observations of PSR 1913+16 with the general relativity
predictions for the first three post-Keplerian parameters, one obtains both a
measurement of the two masses, and a test of the theory, since the system is
overdetermined.  The results are
\begin{equation}
m_1 = 1.4414 \pm 0.0002 M_\odot \,,    m_2 = 1.3867 \pm 0.0002 M_\odot
\,, 
\end{equation}
\begin{equation}
{\dot P_b^{\rm GR}} / {\dot P_b^{\rm OBS}} = 1.0013 \pm 0.0021 \,.
\end{equation}
The concordance among the three constraints on the two masses is shown in Fig.\ \ref{Fig1}.
The accuracy in measuring the relativistic damping of the orbital period
is now limited by uncertainties in our knowledge of the relative
acceleration between the solar system and the binary system as a result of
galactic differential rotation.  In the recently discovered ``double pulsar'' J 0737-3039, all five post-Keplerian parameters are measured, together with the mass ratio $m_1/m_2$ derived directly from the ability to observe the motion of both pulsars.  All six constraints on the masses overlap one another on the $m_1 - m_2$ plane, again consistently with general relativity.

However, there is something potentially wrong here.  All the post-Keplerian effects discussed above are calculated using post-Newtonian theory.  Yet
the neutron stars that compose these systems have {\em very} strong
internal gravity.  This gravitational binding energy reduces the total
mass of each body by 10 to 20 percent  compared to the total rest mass of its constituent particles.  By contrast, the orbital energy is only
$10^{-6}$ of the mass-energy of the system.  Since general relativity is a non-linear theory, surely there is some mixing between the strong internal gravity and the weak interbody gravity.  So how can post-Newtonian theory possibly give valid predictions for such systems?

The reason is a remarkable property of general relativity called the Strong
Equivalence Principle (SEP).  A consequence of this principle is that the
internal structure of a body is ``effaced'', so that the orbital motion and
gravitational radiation emitted by a system of well separated bodies depend {\em only} on the total mass of each body, and not on its internal structure, apart from standard tidal and spin-coupling effects.  In other words, the motion of a normal star or a neutron star or a black hole depends on the body's total mass, and not on the strength of its internal gravitational fields.  This behavior was already implicit in the work of Einstein, Infeld and Hoffman, where only the exterior nearby field of each body was needed, and has been verified theoretically to at least second post-Newtonian order by more modern methods.

By contrast, in alternative theories of gravity, SEP is not valid in
general, and internal-structure effects can lead to significantly
different
behavior, such as the Nordtvedt effect, a possible difference in acceleration of the Earth compared to the Moon in the solar gravitational field, or such as the emission of dipole gravitational radiation from systems of bodies with dissimilar internal structure. 
The close agreement of binary-pulsar data with the predictions of general relativity constitutes a kind of ``null'' test of the effacement of strong-field effects in that theory.  It also constitutes a verification of the unreasonable effectiveness of post-Newtonian theory in this class of strong-gravity systems.

\section{Gravitational waves and inspiralling compact binaries}
\label{sec:bbh}

Possibly the most remarkable example of the unreasonable effectiveness of post-Newtonian theory is that of the inspiral and merger of binary systems of compact objects such as neutron stars and black holes.  The decay of the orbit of a compact binary system through gravitational-wave emission will ultimately bring the two bodies together in a final and catastrophic inspiral, followed by a merger and the likely formation of a terminal black hole.  This process will emit a characteristic gravitational-wave signal with rising frequency and amplitude (often called a ``chirp'') that should be detectable by the world-wide network of ground-based laser-interferometric observatories that is expected to be once again operational following major upgrades, by 2015~\cite{ligo,virgo}.  In the low-frequency end of the gravitational-wave spectrum, the merger of supermassive black holes in the centers of galaxies will be detectable from cosmological distances by the proposed space-based interferometer LISA, currently being planned for a launch after 2020~\cite{lisa}, and by next-generation arrays of radio telescopes doing pulsar timing~\cite{pta}.

The most effective technique for detecting potential binary inspiral gravitational wave signals embedded in the noisy output of these interferometers is the method of {\em matched filtering}, whereby a theoretically generated gravitational-wave signal appropriate for a given source is cross correlated against the output of the detector.  Since the noise is a random process, such a cross-correlation will yield a positive signature if there is a signal that precisely matches the template over the hundreds to thousands of cycles of signal that are expected to lie within the detectable band, even if the signal is formally weaker than the noise.  With a bank of template waveforms that depend on the source parameters such as the two masses, spins, sky location, orbital eccentricity, etc, it will be possible both to detect signals and to measure the properties of the source~\cite{3min}.  To be most effective, this method requires very accurate theoretical templates.  For the inspiral part of the signal, these templates have been calculated by many groups using post-Newtonian theory, with equations of motion (\ref{EOM}) and gravitational waveforms (\ref{waveform}) calculated to 3.5PN order beyond the leading terms~\cite{BlanchetLRR}.

Eventually, however, the inspiral will reach the state where the orbital velocities are high and the gravitational fields are strong, so that the post-Newtonian approximation is no longer valid.  Given our lack of knowledge of the convergence properties of the approximation, it is not known {\em a priori} where this should occur.  

Meanwhile, it has become increasingly clear that the signal from the final few inspiral orbits, from the merger of the two bodies, and even from the final vibrations of the newly formed black hole will make important contributions to the detectability of the waves by the interferometers.  Luckily, many years after its primitive beginnings in the 1970s, numerical relativity finally reached a stage, following critical breakthroughs in 2005~\cite{pretorius}, where researchers could reliably and robustly simulate that final part of the inspiral process.

It therefore came as a complete surprise, when gravitational waveforms from post-Newtonian theory were compared with numerical relativity waveforms for those final orbits, to discover that the agreement was unreasonably good.  The amplitudes and phases of the waves calculated by the two methods agreed remarkably well cycle by cycle over many cycles, and this was in a regime where $Gm/rc^2 \sim 0.2$ and $v/c \sim 0.4$, where one had no right to expect the post-Newtonian approximation to be valid, even with many high-order correction terms in the PN formulae~\cite{baker,boyle,hannam}.  

This unreasonable agreement was crucial, because it permitted the development of techniques for ``stitching together'' post-Newtonian and numerical relativity waveforms to obtain templates that are accurate and valid over the entire inspiral and merger process.  Selecting the best stitching method involves taking into account the noise characteristics of the interferometers whose data is to be analysed, and to find stitchings that optimize all the data analysis protocols, such as false alarm thresholds,  detection confidence criteria, and so on, that are part and parcel of all signal detection strategies.  Because the post-Newtonian waveforms are analytic expansions, they can be resummed using such tricks as Pad\'e approximants to suggest alternative ways to match numerical results.   This ongoing work involves a unique collaboration among post-Newtonian theorists, numerical relativists and interferometer data analysts (see, e.g.~\cite{buonanno2}), but it would have been moot had not post-Newtonian theory been so effective in overlapping with numerical relativity in the strong-gravity high-speed regime.

Another example of the unreasonable effectiveness of post-Newtonian theory relates to the ``kick'' given to a black hole formed from the merger of two compact objects.  In contrast to the erroneous 1937 claim by Levi-Civita, this is a real effect.  If a system emits gravitational waves anisotropically, then the waves carry linear momentum away in addition to energy and angular momentum, and, by virtue of the overall conservation of momentum, the source must recoil in the opposite direction.  It turns out that the gravitational-wave recoil imparted to a final black hole could have important astrophysical consequences, especially for the mergers of supermassive black holes, possibly ejecting the black hole from the host galaxy, or displacing it sufficiently from the center to cause interactions with surrounding gas or stars, thus generating an electromagnetic counterpart signal~\cite{merritt}.

Using formulae for the radiated momentum flux valid to 2PN order, Blanchet {\em et al.}~\cite{bqw} calculated the kick imparted to a black hole from the merger of two non-spinning black holes, as a function of their mass difference (for equal masses, the effect vanishes by symmetry).  The resulting kick velocity turned out to be in remarkable agreement with kicks determined subsequently using numerical relativity, but only up to the point where the two black holes were about to merge.  The agreement was surprising, since the dominant contribution to the recoil comes from a regime where the post-Newtonian approximation should have failed.   

\begin{figure}[t]
\begin{center}
\includegraphics[width=8cm]{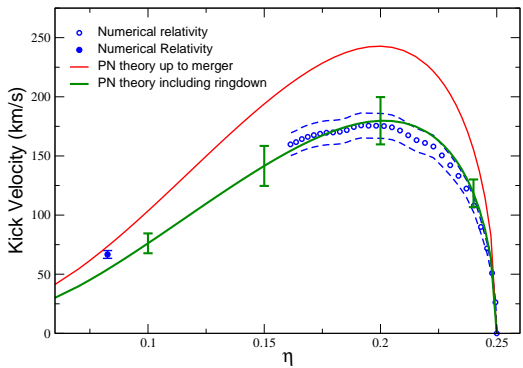}
\caption{Kick velocity vs. reduced mass parameter $\eta$ for non-spinning black hole mergers.  The kick vanishes for equal masses ($\eta=0.25$).  Dashed lines indicate estimated uncertainties in numerical relativity kicks, while error bars indicate estimated uncertainties in the PN-ringdown kicks. }
\label{Fig2}
\end{center}
\end{figure}

The numerical simulations also showed that, following the merger, there was a small ``anti-kick'', reducing the final recoil velocity by around 30 percent~\cite{bakerkick}.  Le Tiec {\em et al.}~\cite{lbw} then used a hybrid calculation that combined formulae for the metric surrounding two closely spaced black holes accurate to 2PN order as initial data, combined with the method of black hole perturbation theory, to study whether the anti-kick was produced by the linear momentum radiated during the ringdown phase of the vibrating final black hole.  The analysis found that the combination of the 2PN inspiral kick plus the ringdown kick was in agreement with the kicks obtained by numerical relativity (Fig. \ref{Fig2}).

A final instance of the effectiveness of post-Newtonian theory came from analyses of the initial configurations used in numerical relativity to study compact binary inspiral.  
Because gravitational radiation tends to circularize binary orbits over time, it is natural to assume that at late times, the inspiralling binary is in a ``quasicircular'' orbit, that is an orbit that is circular, apart from the slow shrinkage due to gravitational-wave damping.  Using numerical relativity it was possible to solve the so-called initial value equations of Einstein's theory for such quasicircular orbits for a variety of systems, including double black holes, double neutron stars, or mixed systems, with and without spin.  These solutions yielded a set of accurate values for the orbital energy $E$ and angular momentum $J$ as a function of the orbital angular velocity 
$\Omega$.  In Newtonian gravity (which would apply to widely separated binaries), these variables would be given by $E = - \eta m (Gm\Omega)^{2/3}/2$ and $J= G\eta m^2 (Gm\Omega)^{-1/3}$.   But the systems in these simulations were highly relativistic, corresponding to $Gm/rc^2 \sim 0.1$ and $v/c \sim 0.3$.  Using post-Newtonian expressions for $E(\Omega)$ and $J(\Omega)$ valid to 3PN order, it was found that the agreement between the PN and numerical results was remarkably good, at the level of several percent in most cases~\cite{blanchet02,morawill1}.   In fact it was suggested that some of the systematic differences between the PN and the numerical results could be explained if the numerical initial configurations actually corresponded to slightly eccentric orbits~\cite{morawill1,morawill2,biw}.  This would change the relation between $E$, $J$ and $\Omega$.  Since one is in a highly relativistic regime, it is not obvious in solving Einstein's initial value equations numerically how to choose the initial separation and initial angular velocity so as to guarantee a circular orbit initially.  Subsequently it was discovered that the numerical evolution of such initial orbits forward in time {\em did} indeed display small amounts of eccentricity.  Numerical methods were then developed to fine-tune the initial configurations to ensure the desired amount of initial eccentricity.

In a related development, Favata recently pointed out the remarkable effectiveness of the PN approximation for determining the final stable circular orbit of two black holes~\cite{favata}.

\section{Concluding remarks}
\label{sec:conclusion}

Wigner remarked that the effectiveness of mathematics in the natural sciences was mysterious.   The unreasonable effectiveness of the post-Newtonian approximation in gravitational physics is no less mysterious.  There is no obvious reason to expect PN theory to account so well for the late stage of inspiral and merger of two black holes.   The Strong Equivalence Principle of general relativity undoubtedly plays a role, by making the internal structure of the bodies irrelevant until they begin to distort one another tidally.  But it does not explain why PN waveforms should agree so well with numerical waveforms when the orbital velocities are almost half the speed of light, or why recoil velocities calculated using PN methods should agree so well with those from numerical methods.  Our colleague Robert V. Wagoner once speculated during the 1970s that, because the gravitational redshift effect makes processes near black holes appear slower and ``weaker'' from the point of view of external observers, the PN approximation should somehow work better than expected, even under such extreme conditions.  Because of the redshift effect, ``strong'' gravity is not as strong as one might think.  But nobody has been able to translate Wagoner's musing into anything quantitative or predictive.  And yet the unreasonable effectiveness of post-Newtonian theory will likely be an important factor in the anticipated first detection of gravitational waves. 





\begin{acknowledgments}
This work was supported in part by the National Science Foundation, grant Nos. PHY06--52448 and PHY09--65133, and the Centre National de la Recherche Scientifique, Programme Internationale de la Coop\'eration Scientifique (CNRS-PICS), Grant No. 4396.  This paper is based in part on a talk given at the Subrahmanyan Chandrasekhar Centennial Symposium at the University of Chicago, October 15 -- 17, 2010.   We are grateful to the Gravitation and Cosmology Group (GReCo) of the Institut d'Astrophysique de Paris for their hospitality during the preparation of this paper, and to Eric Poisson for a critical reading of the manuscript.
\end{acknowledgments}





\end{article}


\begin{table}[t]
\caption{Current Limits on the PPN Parameters}
{\begin{tabular}{c c l}
\noalign{\smallskip}
\hline
Parameter&Limit&Remarks \\
\hline
$\gamma-1$&$2.3 \times 10^{-5}$&Cassini spacecraft tracking\\
&$4 \times 10^{-4}$&VLBI radio deflection \\
$\beta-1$&$3 \times 10^{-3}$&perihelion of Mercury \\
&$2.3 \times 10^{-4}$&no Nordtvedt effect \\
$\xi$&$10^{-3}$&no anomalous Earth tides\\
$\alpha_1$&$10^{-4}$&no anomalies in Lunar, binary\\
&& pulsar orbits\\
$\alpha_2$&$4 \times 10^{-7}$&alignment of Sun
and ecliptic \\
$\alpha_3$&$2 \times 10^{-20}$
&no pulsar ``self'' accelerations \\
$\zeta_1$&$2 \times 10^{-2}$&combined PPN bounds \\
$\zeta_2$&$4 \times 10^{-5}$&no binary ``self''-accelerations \\
$\zeta_3$&$10^{-8}$&no Lunar ``self''-acceleration \\
$\zeta_4$ &-- &not independent \\
\hline
\end{tabular}}
\label{table:bounds}
\end{table}

\end{document}